\documentclass[12pt]{iopart}
\usepackage{graphicx} % Required for the inclusion of images
\usepackage{subfigure} 
\usepackage{amsthm}
\usepackage{amsfonts}
\usepackage{amssymb}
\usepackage{mathrsfs}
\usepackage[utf8]{inputenc}
\usepackage{textcomp}
\usepackage[german, english]{babel}
\usepackage{fixltx2e}
\usepackage{float}
\usepackage{braket}
\usepackage{tikz}
\usepackage{longtable}
\usepackage{url}

\tikzstyle{vertex} = [fill,shape=circle,node distance=80pt]
\tikzstyle{edge} = [opacity=0.5,line cap=round, line join=round, line width= 3pt]
\tikzstyle{elabel} =  [fill,shape=circle,node distance=30pt]
\pgfdeclarelayer{background}
\pgfsetlayers{background,main}

\pgfdeclarelayer{background}
\pgfsetlayers{background,main}

\newtheorem{lemma}{Lemma}
\newtheorem{definition}[lemma]{Definition}
\newtheorem{notation}[lemma]{Notation}

\begin{document}

\title[Error correction with symmetric hypergraph states]{Analysis of quantum error correction with symmetric hypergraph states}

\author{T Wagner, H Kampermann and D Bruß}

\address{Institute for Theoretical Physics 3, Heinrich-Heine University Duesseldorf, Duesseldorf, Germany}
\ead{thomas.wagner@uni-duesseldorf.de}
\vspace{10pt}
\begin{indented}
\item[]September 2017
\end{indented}

\begin{abstract}
Graph states have been used to construct quantum error correction codes for independent errors. Hypergraph states generalize graph states, and symmetric hypergraph states have been shown to allow for the correction of correlated errors. In this paper, it is shown that symmetric hypergraph states are not useful for the correction of independent errors, at least for up to 30 qubits. Furthermore, error correction for error models with protected qubits is explored. A class of known graph codes for this scenario is generalized to hypergraph codes.
\end{abstract}

% Uncomment for PACS numbers
%\pacs{00.00, 20.00, 42.10}
%
% Uncomment for keywords
%\vspace{2pc}
%\noindent{\it Keywords}: XXXXXX, YYYYYYYY, ZZZZZZZZZ
%
% Uncomment for Submitted to journal title message
%\submitto{\JPA}
%
% Uncomment if a separate title page is required
\maketitle
% 
% For two-column output uncomment the next line and choose [10pt] rather than [12pt] in the \documentclass declaration
%\ioptwocol
%

\section{Introduction}

A great challenge in the practical realization of quantum computing is the occurrence of errors on the information that is being processed. Errors on quantum systems can occur independently on each subsystem or in a correlated fashion.
Methods of error correction for independent errors based on graph states have been examined for example in \cite{Schlingemann} and \cite{qditEcc}. In addition, an error model where some qubits are protected from errors was examined and graph codes for this scenario were introduced in \cite{EAQECC}.
In this paper, quantum error correction using \textit{hypergraph states} will be explored. Hypergraph states are a generalization of graph states  introduced in \cite{BP1} and \cite{HypergraphStatesChinese}. Entanglement and nonlocality properties of hypergraph states were explored in \cite{BP2},\cite{ViolationOfLocalRealismInHypergraphStates}, \cite{UnitaryTransformsOfHypergraphStates} and \cite{LocalUnitarySymmetriesOfHypergraphStates}.
Recently, it was shown that \textit{symmetric} hypergraph states can be used to correct certain correlated errors on quantum systems \cite{lyon}. In our paper, the application of symmetric hypergraph states for the correction of independent errors will be explored. It will be shown that symmetric hypergraphs do not lead to good error correction codes for independent error models, at least for up to 30 qubits. Furthermore, graph codes for error models with protected qubits introduced in \cite{EAQECC} are generalized to hypergraph codes.

In the first two sections, the definition and basic properties of hypergraph states as well as the Knill-Laflamme error correction condition will be introduced. Then, the main results about the correction of independent errors with symmetric hypergraphs will be stated. The subcase of graphs will also be explored. Finally, a class of hypergraph codes using protected qubits is constructed. 

\section{Hypergraph States}

\begin{definition}[hypergraph]
Let $V$ be a set and $P(V)$ be the powerset of $V$. A hypergraph $H$ is a pair $H=(V,E)$ where $E \subseteq P(V)$. The elements of V are called nodes or vertices and the elements of E are called edges. Note that a hypergraph may contain the empty edge \{\}.

The cardinality of an edge e is the number of elements in e and is denoted $|e|$. 
An edge with cardinality j is called a j-edge.
A hypergraph where all edges have cardinality 2 is called a graph.
\end{definition}

To define hypergraph states, the concept of a generalized controlled $Z$ gate will be needed.

\begin{definition}[generalized controlled Z gate]
Let $n,k \in \mathbb{N}$ with $k \leq n$. Let $\mathscr{H} = \mathscr{H}_2^{\otimes n}$. Let $e = \{i_1, ... ,i_k\} \subseteq \{1,...,n\}$. Let $j = j_1...j_n$ be a bitstring of size n. The generalized controlled Z gate $C_e = C_{i_1, ... ,i_k}: \mathscr{H} \rightarrow \mathscr{H}$ is the linear operator given by its effect on the computational basis state $\ket{j}$:
\begin{equation}
C_{i_1, ... ,i_k}\ket{j} = (-1)^{j_{i_1} \cdot j_{i_2} \cdot ... \cdot j_{i_k}}  \ket{j}
\end{equation}
\end{definition} 

That means $C_{i_1, ... ,i_k}$ will flip the sign of $\ket{j}$ if $j_{i_1} = j_{i_2} = ... = j_{i_k} = 1$ and act as the identity otherwise. Per definition $C_{\{\}} = -1$, so the state is only changed by a global phase.

For a hypergraph $H$, a corresponding quantum state can be defined as follows \cite{BP2}:

\begin{definition}[hypergraph state]
Let $H = (V,E)$ be a hypergraph with n nodes. The corresponding hypergraph state $\ket{H}$ is given by:
\begin{equation}
\ket{H} = (\Pi_{e \in E} C_e)\ket{+}^{\otimes n}
\end{equation}
\end{definition}

In the following, $X_i,Z_i$ and $Y_i$ will be used to denote the Pauli $X,Z$ or $Y$ operator acting only on qubit $i$, tensored with the identity on all other qubits.

With each hypergraph state a \textit{hypergraph basis} can be associated:

\begin{definition}[hypergraph basis]
\label{hypergraph basis}
Let $H = (V,E)$ be a hypergraph with n nodes. Let $\ket{H}$ be the corresponding hypergraph state.

The hypergraph basis associated with the hypergraph state $\ket{H}$ is the set
\begin{equation}
B_H = \{(\Pi_{i \in I}Z_i) \ket{H}| I \subseteq \{1,...,n\}\}
\end{equation}
\end{definition}

This is an orthonormal basis of the space $\mathscr{H}_2^{\otimes n}$ \cite{BP2}. 

The Pauli-Z operator always maps a hypergraph state to an orthogonal hypergraph state. The effect of the Pauli-X operator is described by the following lemma \cite{BP2}:

\begin{lemma} \cite{BP2}
\label{XeffectLemma}
Let $H = (V,E)$ be a hypergraph with n nodes. Let $\ket{H}$ be the corresponding hypergraph state. Let $E(i) = \{e \in E | i \in e\}$.Then:
\begin{equation}
\label{Xeffect}
X_i \ket{H} = \Pi_{e \in E(i)} C_{e \setminus \{i\}} \ket{H}
\end{equation}
\end{lemma}

Finally, the concept of a symmetric hypergraph will be defined:
\begin{definition}[symmetric hypergraph]
Let $H = (V,E)$ be a hypergraph with n nodes.

H is called \textbf{symmetric} if and only if H does not change under any permutation of the nodes. Equivalently, H is symmetric if and only if, for any possible cardinality of edges $m < n$, H either contains no edges of cardinality m or all possible edges of cardinality m.
\end{definition}

It was shown in \cite{lyon}, that for a symmetric hypergraph state $\ket{H}$, the coefficient of the computational basis state $\ket{s}$ (more formally: the scalar product $\braket{H|s}$) depends only on the weight of $s$, that is, the number of ones in the bit-string $s$.
This allows us to introduce a simple notation (this notation is different from the one used in \cite{lyon}):

\begin{notation}
\label{NotationSymmetricHypergraphTupleWeights}
Let $n,k \in \mathbb{N}$ with $k \leq n$. Let $M = \{m_1,...,m_k\} \subseteq \{ 1,...,n \}$. Then, the symmetric hypergraph state on n qubits, where all computational basis states with a weight in M have a negative coefficient and all other computational basis states have a positive coefficient, is denoted $\ket{H^{m_1,...,m_k}_n}$.

That is:
\begin{equation}
\ket{H^{m_1,...,m_k}_{n}} = \frac{1}{\sqrt{2^n}} \sum_{s \in \mathbb{(Z}_2)^n} a_s \ket{s}
\end{equation}
where 
\begin{equation}
a_s = \cases{
+1,|s| \notin M \\
-1,|s| \in M 
}
\end{equation}   
\end{notation}

\section{Knill-Laflamme Condition}

For protection against errors, qubits can be encoded into a subspace $C$ of a larger Hilbert space $\mathscr{H}$. This subspace $C$ is then called a quantum code. A quantum code $C$ is said to correct the quantum operation $\varepsilon: \mathscr{H} \rightarrow \mathscr{H}$ described by operational elements {$E_k$} if a quantum operation $R: \mathscr{H} \rightarrow \mathscr{H}$ exists with the property that $R \circ \varepsilon   (\ket{\psi}\bra{\psi}) \propto \ket{\psi}\bra{\psi}$ for all $\ket{\psi} \in C$.
In order to examine the error correction capabilities of codes, the following condition will be used \cite{KL97}: 

\begin{lemma}[KL-condition] \cite{KL97}
\label{KLCond}
Let $C \subset \mathscr{H}$ be a quantum code and P the projector onto C. Let $B$ be an orthonormal basis of C. Let $\varepsilon$ be a quantum operation described by a set $\mathscr{E}$ of operational elements. Then, a recovery operation $R$ correcting $\varepsilon$ exists if and only if for all $\ket{b_l},\ket{b_m} \in B$ and $E_i,E_j \in \mathscr{E}$ the following holds:

\begin{equation}
\label{eccdiag}
\braket{b_l|E_i^{\dagger}E_j|b_m} = \cases{
0, \textrm{ if }  l \neq m \\
\alpha_{ij}, \textrm{ if }  l=m
}
\end{equation}
for some complex numbers $\alpha_{ij}$, only depending on $i$ and $j$.
\end{lemma}

The condition for $l = m$ will usually be referred to as the diagonal condition, the other one will be referred to as the off-diagonal condition. 
It is an important result that if a code is capable of correcting the errors described by the four Pauli-operators  $I_i,X_i,Y_i,Z_i$ on a given qubit $i$, 
then it is capable of correcting arbitrary errors on that qubit \cite{NC}. 

\begin{definition}[distance]
Let $d\in \mathbb{N}$.
A code C is said to have a distance $d$ if the KL-condition is fulfilled for any operator $E=E_i^{\dagger}E_j$ that acts non-trivially on strictly less than $d$ qubits.
\end{definition}
A code with distance $d$ can correct all errors on up to $(d-1)/2$ qubits, so it is useful for the correction of independent errors.

\section{Results}

\subsection{Symmetric Hypergraph Codes: Construction and Properties}

In this section some hypergraph codes will be considered and it will be investigated which codes can have a distance of at least 2. The main result will be a classification of symmetric hypergraph codes with distance at least 2 in terms of binomial coefficients, and a necessary condition for distance at least 3.

Two main ideas are used:
The first idea is that the scalar product of two hypergraph states depends not on the states themselves, but only on the difference in edges between the states. This is important because only scalar products between states are relevant for the KL-condition.
The second idea is that, to fulfill the KL-condition, it is essentially sufficient to find states that get mapped to orthogonal states by the Pauli $X$ operator, as the effect of the Pauli $Z$ operator on hypergraph states is relatively simple. A Pauli $Z$ operator maps hypergraph states to states in their hypergraph basis (Definition \ref{hypergraph basis}), which are always orthogonal to the original state. 
Motivated by this, the following basic construction will be explored:
Find a hypergraph state $\ket{D}$ that corresponds to the difference in edges between two orthogonal states (such states will be called balanced states, see Definition \ref{balanced}). Then, construct a state $\ket{H}$ so that the difference between the hypergraphs $H$ and $X_1H$ is $D$. That error correction works, not only on the first but on any qubit, can be guaranteed by choosing $H$ as a symmetric hypergraph, so all qubits can be treated in the same way. The state $\ket{H}$ will then be used to define a quantum code encoding one qubit as span($\ket{H}$,$\ket{G}$) where $\ket{G}$ is any state in the hypergraph basis of $H$. Which choice of $\ket{G}$ leads to a useful code will be analyzed in Lemma \ref{ecc}.

In the following, these heuristics are formalized and a basic error correction condition for this kind of code is derived.
First, some useful notation is introduced:

\begin{notation}[symmetric difference of hypergraph states]
\label{NotationSymmetric}
Let H = (V,E$_H$) and G =(V,E$_G$) be two hypergraphs with n vertices each. Let $\ket{H}$ and $\ket{G}$ be the associated hypergraph states. Then  the hypergraph (V,E$_H$ $\Delta$ E$_G$) will be denoted  H $\Delta$ G. Here, $\Delta$ denotes the symmetric difference between sets: $E_H \Delta E_G = E_H \cup E_G \setminus (E_H \cap E_G)$  The corresponding hypergraph state is then denoted $\ket{H \Delta G}$ or $\ket{H} \Delta \ket{G}$.
\end{notation}

To further investigate the scalar product of hypergraph states the concept of a balanced state will be useful:
\begin{samepage}

\begin{definition}[balanced]
\label{balanced}
Let $\ket{R}$ be a hypergraph state on n qubits. Then $\ket{R}$ is called \textbf{balanced} if and only if $\ket{R}$ contains $2^{(n-1)}$ negative and $2^{(n-1)}$ positive  coefficients in its representation in the computational basis. Equivalently, $\ket{R}$ is balanced if and only if $\ket{R} \bot \ket{+}^{\otimes n} $.
\end{definition} 
\end{samepage}

Note that if $\ket{R} = \ket{+}\ket{R^{\prime}}$ for some state $\ket{R^{\prime}}$ on $n-1$ qubits, then $\ket{R}$ is balanced if and only if $\ket{R^{\prime}}$ is balanced, which holds because of:

$\ket{R} = \frac{1}{\sqrt{2}}( \ket{0}\ket{R^\prime} + \ket{1}\ket{R^\prime})$

Both summands have the same number of negative coefficients, so they must have $2^{(n-2)}$ negative coefficients each for $\ket{R}$ to have $2^{(n-1)}$ negative coefficients.
\\

Now, the scalar products of hypergraph states will be examined: (See also \cite{ScalarGraph} for the case of graph states)

\begin{lemma}
\label{ortho}
Let H = (V,E$_H$) and G =(V,E$_G$) be two hypergraphs with n vertices each. Let $\ket{H}$ and $\ket{G}$ be the associated hypergraph states.

Then $\braket{H|G} = \bra{+}^{\otimes n} \ket{H \Delta G}$.

That is, the scalar product of two hypergraph states depends only on the symmetric difference between their sets of edges.

Specifically, $\ket{H} \bot \ket{G}$ if and only if $\ket{H \Delta G}$ is balanced.
\end{lemma}

\textbf{Proof:}
\begin{eqnarray}
\braket{H|G}& = \bra{+}^{\otimes n}\Pi_{e \in E_H} C_e \Pi_{e \in E_G} C_e \ket{+}^{\otimes n} = \bra{+}^{\otimes n}\Pi_{e \in (E_H \Delta E_G)} C_e \ket{+}^{\otimes n}\nonumber \\ &=  \bra{+}^{\otimes n}\ket{H \Delta G}
\end{eqnarray}
The first equality follows directly from the definition of a hypergraph state. The second equality follows because the generalized controlled $Z$ gates are self-inverse and commute.
\begin{flushright}
$\square$
\end{flushright}

In the following, symmetric hypergraph codes $C$ in the sense of the following definition will be considered:

\begin{definition}[symmetric hypergraph code]
\label{DefinitionSymmetricHypergraphCode}
Let H = (V,E) be a symmetric hypergraph with n vertices. Let  $I \subseteq \{1,.,n\}$. 
The \textbf{symmetric hypergraph code} C associated with H and I is:
$C = span(\ket{H},\ket{H^\prime})$, where $\ket{H^\prime} = \Pi_{i \in I}Z_i\ket{H}$.

Furthermore, a code C on n qubits is called a \textbf{symmetric hypergraph code} if there exists a symmetric hypergraph H and a set $I \subseteq \{1,.,n\}$ such that C can be written in the form above.
\end{definition}

Because of the symmetry of the states, in the following just the cardinality $l = |I|$ of $I$ will be given and without loss of generality it is assumed that $I = \{1,..,l\}$. Furthermore, it is assumed without loss of generality that $H$ contains no edges of size 1, and that $H$ does not contain the empty edge. (Those edges can always be eliminated by a local basis change)

A very useful concept in the calculations will be the X-difference hypergraph, which will often be used to express scalar products of certain hypergraph states:

\begin{definition}[X-difference hypergraph]
\label{DefinitionX-differenceHypergaph}
Let H be a symmetric hypergraph with n vertices. Let $i \in \{1,...,n\}$.
Let $E(i) = \{e\in E| i \in e\}$ be the set of edges in E containing the vertex i.

We define the hypergraph $D_{\overline{i}}$ with n-1 vertices by removing vertex $i$ and all edges containing this vertex:

$D_{\overline{i}} = (V\setminus \{i\},E_D)$, where $E_D = \{e\setminus \{i\}|e \in E(i)\}$

Note that because H was symmetric, $D_{\overline{i}}$ is also symmetric and actually independent of the choice of i.
We call $D := D_{\overline{i}}$ the \textbf{X-difference hypergraph} associated with H.

Then, the following equation holds:

\begin{equation}
\ket{H} \Delta (X_j\ket{H}) = \ket{+}_j\ket{D}_{\overline{j}}
\label{Xdifference}
\end{equation}
\end{definition}

\textbf{Proof of equation (\ref{Xdifference}):}

With the notation above:

\begin{eqnarray}
\ket{H} \Delta (X_j\ket{H}) &= \ket{H} \Delta (\Pi_{e \in E(j)} C_{e \setminus \{j\}} \ket{H}) \nonumber \\ &=\Pi_{e \in E(j)} C_{e \setminus \{j\}} \ket{+}^{\otimes n} = \ket{+}_j \Pi_{e \in E(j)} C_{e\setminus \{j\}} \ket{+}^{\otimes n-1} \nonumber \\ &= \ket{+}_j\ket{D}_{\overline{j}}
\end{eqnarray}

Here, the first equality follows from (\ref{Xeffect}) and the last equality is by definiton of $D$.

\begin{flushright}
$\square$
\end{flushright}

With this definition, the construction mentioned at the beginning of this section can be formalized:

\begin{lemma}[construction of codes]
\label{tupleCorrespondence}

Let $C$ be a symmetric hypergraph code on n qubits associated with a symmetric hypergraph H = (V,E) and a set $I \subseteq \{1,.,n\}$.(Definition \ref{DefinitionSymmetricHypergraphCode}).
Let D be the X-difference hypergraph associated with H. (Definition \ref{DefinitionX-differenceHypergaph})
The code C can be given by the hypergraph D and the number $l = |I|$.
Because $\ket{D}$ is a symmetric hypergraph state, there exist natural numbers
$m_1,...,m_k \in \{1,..,n-1\}$ such that $\ket{D} =\ket{H^{m_1,..,m_k}_{n-1}}$. (See Notation \ref{NotationSymmetricHypergraphTupleWeights})
In this way (on a given number of qubits) a \textbf{unique} tuple $(\{m_1,..,m_k\},l)$ can be associated with every symmetric hypergraph code C. 
Conversely, every tuple $(\{m_1,..,m_k\},l)$ is associated with a unique code. (On a given number of qubits)
\end{lemma}
\textbf{Proof}:
The code can be constructed from the tuple in the following way:

Write down the coefficients of the $n-1$ qubit state $\ket{D}_{\overline{1}} = \ket{H^{m_1,..,m_k}_{n-1}}$ in the computational basis. From this, the edges of $D = (V \setminus \{1\}, E_D)$ can be constructed by the algorithm given in \cite{BP1}. Now, introduce a new vertex 1 to $D$ and replace all edges $e$ of $D$ by the edge $e \cup \{1\}$, that is define the set $E_D^{\prime} = \{e \cup \{1\}| e \in E_D\}$. Now all edges containing the vertex 1 are determined. Because the constructed hypergraph will be symmetric this already uniquely determines the constructed state. That is, define $H = (V,E)$ as the unique symmetric hypergraph not containing the empty edge that fulfills $E(1) = E_D^{\prime}$. Then define the Code $C$ as $span(\ket{H},\ket{H^\prime})$ where $\ket{H^\prime} = \Pi_{i \in I}Z_i\ket{H}$ and $I = \{1,..,l\}$. 

\begin{flushright}
$\square$
\end{flushright}

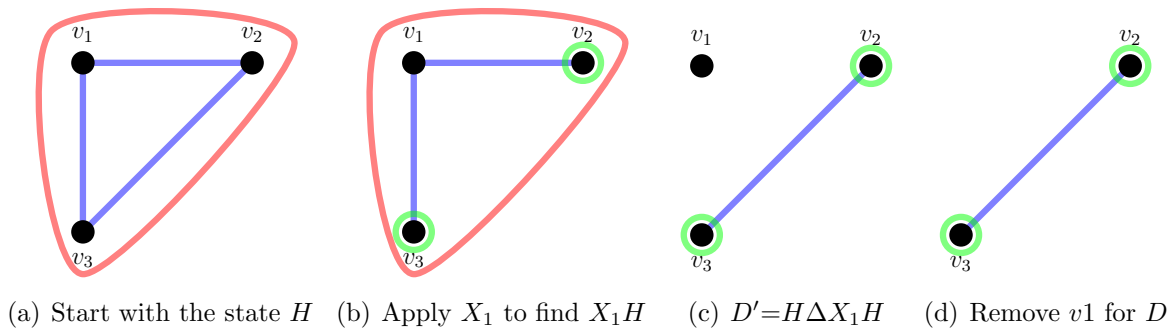
\begin{figure}[H]
\subfigure[Start with the state $H$]
{
\scalebox{0.8}
{
\begin{tikzpicture}
\node[vertex,label=above:\(v_1\)] (v1) {};
\node[vertex,right of=v1,label=above:\(v_2\)] (v2) {};
\node[vertex,below of=v1,label=below:\(v_3\)] (v3) {};

\begin{pgfonlayer}{background}
\draw [edge,red] plot [smooth cycle] coordinates {([shift=({-15pt,+15pt})] v1) ([shift=({+20pt,+10pt})]v2) ([shift=({0pt,-20pt})]v3)};
\draw[edge,color=blue](v1)--(v2);
\draw[edge,color=blue](v1)--(v3);
\draw[edge,color=blue](v2)--(v3);
\end{pgfonlayer}
\end{tikzpicture}
}
}
\subfigure[Apply $X_1$ to find $X_1H$]
{

\scalebox{0.8}
{
\begin{tikzpicture}
\node[vertex,label=above:\(v_1\)] (v1) {};
\node[vertex,right of=v1,label=above:\(v_2\)] (v2) {};
\node[vertex,below of=v1,label=below:\(v_3\)] (v3) {};

\begin{pgfonlayer}{background}
\draw [edge,red] plot [smooth cycle] coordinates {([shift=({-15pt,+15pt})] v1) ([shift=({+20pt,+10pt})]v2) ([shift=({0pt,-20pt})]v3)};
\draw[edge,color=blue](v1)--(v2);
\draw[edge,color=blue](v1)--(v3);
\draw[edge,color=green](v3) circle (0.3cm);
\draw[edge,color=green](v2) circle (0.3cm);
\end{pgfonlayer}
\end{tikzpicture}
}
}
\subfigure[$D^\prime$=$H\Delta X_1H$]
{
\scalebox{0.8}
{
\begin{tikzpicture}
\node[vertex,label=above:\(v_1\)] (v1) {};
\node[vertex,right of=v1,label=above:\(v_2\)] (v2) {};
\node[vertex,below of=v1,label=below:\(v_3\)] (v3) {};

\begin{pgfonlayer}{background}
\draw[edge,color=blue](v2)--(v3);
\draw[edge,color=green](v3) circle (0.3cm);
\draw[edge,color=green](v2) circle (0.3cm);
\end{pgfonlayer}
\end{tikzpicture}
}
}
\subfigure[Remove $v1$ for $D$]
{
\scalebox{0.8}
{
\begin{tikzpicture}
\node[vertex,right of=v1,label=above:\(v_2\)] (v2) {};
\node[vertex,below of=v1,label=below:\(v_3\)] (v3) {};

\begin{pgfonlayer}{background}
\draw[edge,color=blue](v2)--(v3);
\draw[edge,color=green](v3) circle (0.3cm);
\draw[edge,color=green](v2) circle (0.3cm);
\end{pgfonlayer}
\end{tikzpicture}
}
}
\caption{Construction of $D$ (see Definition \ref{DefinitionX-differenceHypergaph} and Lemma \ref{tupleCorrespondence}) from a symmetric hypergrah code}
\label{TupleConstructionFigure}
\end{figure}

Example:

Figure \ref{TupleConstructionFigure} illustrates how to construct $D_{\overline{1}}$ from a given hypergraph $H$.

In this example, the state corresponding to $D$ is $\ket{D}_{\overline{1}} = \frac{1}{2}(\ket{00}-\ket{01}-\ket{10}-\ket{11})$. Note that $D$ is symmetric. Therefore this state corresponds to the tuple $(1,2)$. A code C on 3 qubits could for example be defined as $(\{1,2\},1)$. Then $C = span(\ket{H},Z_1\ket{H})$. 
Conversely, the state $H$ can be constructed from $D$ by going backwards through the diagram in Figure \ref{TupleConstructionFigure}, following the steps explained in Lemma \ref{tupleCorrespondence}.

Now, the KL-condition (Lemma \ref{KLCond}) will be used to examine this kind of code. The case $l = 0 $ is trivial, and in the case $l = 1$ single $Z$ errors cannot be corrected. The other cases are described in the following lemma:

\begin{lemma}
\label{ecc}
Let C = span($\ket{H},\ket{H^\prime}$) be a symmetric hypergraph code (Definition \ref{DefinitionSymmetricHypergraphCode}) on n qubits, and let (\{$m_1,...,m_k$\},l) be the tuple describing this code, where $l>1$. (Therefore: $\ket{H^\prime} = \Pi_{j=1}^l Z_j \ket{H}$.) Let $\ket{D} = \ket{H^{m_1,..,m_k}_{n-1}}$.
Then C has a distance of at least 2 if and only if all of the following conditions hold:

\begin{enumerate}
\item $\ket{D}$ is balanced.
\item $\Pi_{i=1}^l Z_i\ket{D}$ is balanced.
\item $\Pi_{i=1}^{l-1}Z_i\ket{D}$ is balanced.
\end{enumerate}
\end{lemma}
\textbf{Proof:} See appendix.

Now Lemma \ref{ecc} will be rewritten in terms of binomial coefficients. 
But first, a preliminary remark regarding the relation of Lemma \ref{ecc} and Lemma \ref{eccBinomial}:
In the proof of Lemma \ref{eccBinomial} it will become clear that the conditions (i) in lemma \ref{ecc} and \ref{eccBinomial} correspond directly to each other. However, for the correspondence between the conditions (ii), and the correspondence between the conditions (iii), it is needed that condition (i) is fulfilled. Therefore the whole set of conditions is equivalent, but the individual conditions are not.

For convenience of notation, in the following lemma the letter $n$ is used to denote the number of qubits of $D$, not the number of qubits of $H$ as in the previous lemmas.

\begin{lemma}
\label{eccBinomial}
Let C be the code described by the tuple $(\{m_1,...,m_k\},l)$ on \textbf{n+1} qubits.
Then C has a distance of at least 2 if and only if all of the following criteria hold:

\begin{enumerate}
\item $\sum_{m \in M} {n \choose m} = 2^{n-1}$
\item $\sum_{m \in M}\sum_{j \ \rm{odd}, j \leq l, j \leq m } {l \choose j}{n-l \choose m-j} = 2^{n-2}$
\item $\sum_{m \in M}\sum_{j \ \rm{odd}, j \leq l-1, j \leq m} {l-1 \choose j}{n-(l-1) \choose m-j} = 2^{n-2}$
\end{enumerate}
\end{lemma}
\textbf{Proof:} See appendix.

With this lemma, it can be investigated which codes have a distance of at least 2.
In addition, it also gives a necessary condition for distance at least 3.
If a symmetric hypergraph code with number of $Z$ gates $l$ has distance at least 3, the codes with $l+1$ and $l-1$ $Z$ gates must have distance at least 2. Even more general, the following lemma holds:

\begin{samepage}
\begin{lemma}
\label{dist3condition}
Let C be the symmetric hypergraph code described by the tuple (\{$m_1,...,m_k$\},l) on n qubits, where $1<l<n$.
\nopagebreak
If C has a distance of at least d, then the codes (($m_1,...,m_k$),l+1) and
\nopagebreak
($m_1,...,m_k$),l-1) have a distance of at least d-1.
\end{lemma}
\end{samepage}

\textbf{Proof}:

By the definition of a symmetric hypergraph code, there exists a state $\ket{H}$ such that: $C = span(\ket{H},\ket{H^\prime})$ where $\ket{H^\prime} = \Pi_{j=1}^l Z_j \ket{H}$.

First, it is shown that the code $N$ described by the tuple $(\{m_1,...,m_k\},l+1)$ has distance at least $d-1$:

From the definition of the codes it is clear that:

$N = span(\ket{H},Z_{l+1}\ket{H^\prime})$. This simply means that $N$ has one more $Z$ gate for the second codeword. Now the KL-condition (Lemma \ref{KLCond}) can be shown:

Let $E_i$ be a Pauli $X$,$Z$ or $XZ$ operator acting on qubit $i$.

Then:
\begin{equation}
\braket{H|E_iZ_{l+1}|H^\prime} = 0
\end{equation}

The equality holds because $C$ has distance at least $d$, so it fulfills the KL-conditions $\braket{H|E|H^\prime} = 0$ for any operator of the form $E = E_iZ_j$ where $j \in \{1,..,n\}$.

This shows the off-diagonal KL-condition for distance at least 2 for the code $N$.

Furthermore:
\begin{equation}
\braket{H|E_i|H} =  \braket{H^\prime|Z_{l+1}E_iZ_{l+1}|H^\prime}
\end{equation}

The equality holds because $C$ has distance at least $d$, so in particular it fulfills the KL-condition $\braket{H|E|H} = \braket{H^\prime|E|H^\prime}$ for the operator $E = Z_{l+1}E_iZ_{l+1}$.

This shows the diagonal KL-condition for $N$.

Therefore it was shown that the code $N$ described by the tuple $(\{m_1,...,m_k\},l+1)$ has distance at least $d-1$.

With the same arguments, using the gate $Z_{l}$ instead of $Z_{l+1}$ it can also be shown that $((m_1,...,m_k),l-1)$ must have distance at least $d-1$.

\begin{flushright}
$\square$
\end{flushright}

\subsection{Computer Search for Symmetric Hypergraph Codes}

The condition given in Lemma \ref{eccBinomial} made it possible to perform a computer search for codes. A systematic search for tuples fulfilling the given conditions was conducted, and a number of codes with distance 2 was found. The search was restricted to actual hypergraph codes, and normal graph codes were filtered out.(Normal graph codes are analyzed in Section \ref{GraphCodesSection}) The smallest actual hypergraph code was found on 8 qubits. The search was performed up to 30 qubits.
The results were filtered for codes that fulfill the \textbf{necessary} condition for distance 3 given in lemma \ref{dist3condition}. Four results were found, all on 30 qubits.
However, using the original KL-condition, it was confirmed that none of these codes actually have a distance of 3. 
Therefore it can be concluded that no code of the given form on less than 30 qubits has a distance 3 or higher. 

The program code is available at \cite{Program1} and
\cite{Program2}. A list of all symmetric hypergraph codes for up to 20 qubits is avaible at \cite{Program2}.

\subsection{The Special Case of Graph Codes}
\label{GraphCodesSection}
In this section, the special case of graph codes will be further examined. The computer results above suggest that it is difficult to find symmetric hypergraph codes with a distance of at least 3. For the special case of graph codes, it can be confirmed analytically that no symmetric graph code can have a distance of 3 or higher.

\begin{lemma}
Let $n \in \mathbb{N}$ and $l \leq n$. Let G be a symmetric graph. Let $\ket{G}$ be the corresponding graph state, and $\ket{G^\prime} = \Pi_{i=1}^lZ_i\ket{H}$. Let $C = span(\ket{G},\ket{G^\prime})$. Then the code C does not have a distance of 3 or higher.
\end{lemma}

\textbf{Proof}:
If $l = n$, then $X_1Z_1 \ket{H} = \ket{H^\prime}$. This is because of equation (\ref{Xeffect}) and the fact that $C_i = Z_i$ for generalized controlled $Z$ gates with only 1 index.
Then, the off-diagonal KL-condition for distance 3 is already violated.
If $l = 0$, then $\ket{G} = \ket{G^\prime}$ and the result is obvious.

For $0 \neq l < n$:
\begin{eqnarray}
\braket{G|Z_1X_1Z_nX_n|G} &= \braket{G|Z_1X_1Z_n   \Pi_{i=1}^{n-1}Z_i|G} \nonumber \\ &= - \braket{G|\Pi_{i=1}^nZ_i\Pi_{i=1}^nZ_i|G} = -1
\end{eqnarray}
The first equality follows from the way $X$ operators operate on graph states (equation (\ref{Xeffect})). The other equalities follow by commuting operators and applying the other $X$ gate.

On the other hand:
\begin{eqnarray}
\braket{G^\prime|Z_1X_1Z_nX_n|G^\prime} &= \braket{G|\Pi_{i=1}^lZ_i Z_1X_1Z_nX_n \Pi_{i=1}^lZ_i|G} \nonumber \\ &= - \braket{G|Z_1X_1Z_nX_n|G}
\end{eqnarray}
The last equality follows because $X_n$ and $\Pi_{i=1}^lZ_i$ commute ($l < n$), but $X_1$ and $\Pi_{i=1}^lZ_i$ anti-commute ($l > 0$).
Therefore the KL-condition is violated for the operator $Z_1X_1Z_nX_n$ and the code does not have a distance of 3 or higher.

\begin{flushright}
$\square$
\end{flushright}

\section{Constructing Hypergraph Codes with Protected Qubits}

In some scenarios there exist specially protected qubits that are not liable to errors, for example in a communication scenario with a preexisting entangled pair of qubits that do not have to be sent through a channel. Another example are quantum computers that use two different kinds of qubits with different error rates. For example, in a register employing electron and nuclear spin, the electron spin decays on the order of $\mu s$, while the nuclear spin decays only on the order of $ms$ \cite{Decay}.
General codes for such a scenario were studied in \cite{EAQECCBrun}.
In \cite{EAQECC} it was examined how this kind of protected qubit can be used to construct efficient error correcting codes using graph states. In particular, codes arising from so-called star graphs were considered.
The principle of such a code is to use a graph consisting of a central vertex that is connected to all other vertices, and no connections between the outer vertices (a star graph). The second codeword is obtained by applying $Z$ gates to all outer vertices. In such a graph, an $X$ error on the central vertex corresponds to $Z$ errors on all outer vertices. Therefore such an error maps codewords to codewords and is not correctable. However, if the central vertex is protected, most errors on the outer vertices can be identified and corrected. 
In this section this notion will be generalized and hypergraph codes with protected qubits will be constructed.

This can be done by starting with a hypergraph $D$ that corresponds to a balanced quantum state $\ket{D}$ and adding a number of outer vertices to $D$. The set of new vertices will be called $A$. The hypergraph $D$ will later play a similar role to the X-difference hypergraph $D$ used in previous proofs.
It is assumed that the original vertices of $D$ correspond to protected qubits. In contrast to the previous section, $D$ is not necessarily symmetric. For each edge $e$ of $D$ and each outer vertex $i$ a new edge $e \cup \{i\}$ is introduced, and the original edge $e$ is removed. Arbitrary edges can be introduced on the original vertices of $D$. The new graph constructed in this way is called $H$.

\begin{lemma}
\label{lemmaGeneralEAECC}
Let $e,n \in \mathbb{N}$.
Let $D = (P,E_D)$ be a hypergraph with e (protected) vertices such that the corresponding quantum state $\ket{D}$ is balanced.  

Let $O = \{I \subset P | \bra{D}\Pi_{i \in I} Z_i \ket{+}^{\otimes e} = 0 \}$. 

(The elements of $O$ correspond to the hypergraph basis states that do not appear in the representation of $\ket{D}$ in the hypergraph basis of $\ket{+}^{\otimes n}$)

Let $H = (A \cup P,E_H)$ be a hypergraph on n + e qubits with corresponding quantum state $\ket{H}$ (A denotes a set of n new vertices)

Suppose that for all $i \in A$ it holds: 
\begin{equation}
X_i \ket{H} = \Pi_{e \in E_D} C_e \ket{H}
\end{equation} 

Choose a set $S \subset O$ such that for all $I,J \in S$: $I \Delta J \in O$.

For each $I \in S$ let $\ket{H_I}$ be the state $\Pi_{i \in I} Z_i \ket{H}$ and  $\ket{H_I^\prime} = \Pi_{i \in A} Z_i \ket{H_I}$

Let $C = span_{I \in S}(\ket{H_I}, \ket{H_I^\prime})$.
Then C corrects the error set $R$ consisting of all operators $E_i,E_j$ acting on A with $weight(E_iE_j^\dagger) < |A|$.
\end{lemma}

\textbf{Proof:} See appendix.

Note that $H$ can be constructed by replacing the edges of $D$ as explained above Lemma \ref{lemmaGeneralEAECC}.

There are two remarks concerning the above lemma:
First, the easiest way to get a large set $O$ (and therefore a large codespace) is to choose $\ket{D} =  \Pi_{i \in {1,..,e}} Z_i \ket{+}^{\otimes e}$. The resulting state $\ket{H}$ will then be a regular graph state. It is not clear whether using actual hypergraph states offers any advantage over this.
Second, if no basis states with $Z$ gates on the outer qubits are used in the code $C$, it is clear from the proof that the code $C$ will actually correct any errors on the outer qubits, leading to stronger error correction capabilities. This is because the error $\Pi_{i \in A} Z_i$ will then no longer map codewords to codewords.

Consequently, using such a code will, in principle, make the outer qubits completely safe. This means that the error rate of the code is determined by the error rate of the protected qubits. However, the actual implementation of the decoding seems problematic, as non-local measurements on both the protected and unsafe qubits need to be implemented.

\subsection{Example of a Hypergraph Code With Protected Qubits}

Here, a simple example of a hypergraph code with two protected and two unsafe qubits will be presented, following the procedure outlined in the preceding section.
The code is constructed by choosing $\ket{D} = Z_1Z_2\ket{+}^{\otimes 2}$, and then adding two qubits to obtain $H$.
The set $O$, introduced in Lemma \ref{lemmaGeneralEAECC}, is then, according to its definition:
\begin{equation}
O = \{ \varnothing , \{1\} , \{2\}  \}
\end{equation}

A possible choice of the set $S$ is:
\begin{equation}
S = \{ \varnothing, \{1\}\}
\end{equation}

Note that the choice: 
\begin{equation}
S = \{ \varnothing, \{1\},\{2\}\}
\end{equation}
is \textbf{not} possible because

$\{1\} \Delta \{2\} = \{1,2\} \notin O$.

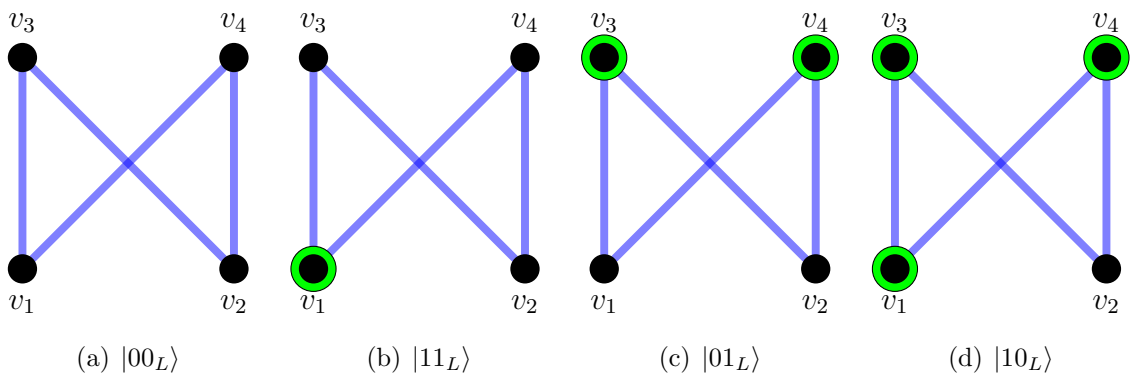
\begin{figure}[H]
\subfigure[$\ket{00_L}$]
{
\begin{tikzpicture}
\node[vertex,label=below:\(v_1\)] (v1) {};
\node[vertex,right of=v1,label=below:\(v_2\)] (v2) {};
\node[vertex,above of=v1,label=above:\(v_3\)] (v3) {};
\node[vertex,right of=v3,label=above:\(v_4\)] (v4) {};

\begin{pgfonlayer}{background}
\draw[edge,color=blue](v1)--(v3);
\draw[edge,color=blue](v2)--(v3);
\draw[edge,color=blue](v1)--(v4);
\draw[edge,color=blue](v2)--(v4);
\end{pgfonlayer}
\end{tikzpicture}
}
\subfigure[$\ket{11_L}$]
{
\begin{tikzpicture}
\node[vertex,label=below:\(v_1\)] (v1) {};
\node[vertex,right of=v1,label=below:\(v_2\)] (v2) {};
\node[vertex,above of=v1,label=above:\(v_3\)] (v3) {};
\node[vertex,right of=v3,label=above:\(v_4\)] (v4) {};

\begin{pgfonlayer}{background}
\draw[edge,color=blue](v1)--(v3);
\draw[edge,color=blue](v2)--(v3);
\draw[edge,color=blue](v1)--(v4);
\draw[edge,color=blue](v2)--(v4);
\filldraw[fill=green](v1) circle (0.3cm);
\end{pgfonlayer}
\end{tikzpicture}
}
\subfigure[$\ket{01_L}$]
{
\begin{tikzpicture}
\node[vertex,label=below:\(v_1\)] (v1) {};
\node[vertex,right of=v1,label=below:\(v_2\)] (v2) {};
\node[vertex,above of=v1,label=above:\(v_3\)] (v3) {};
\node[vertex,right of=v3,label=above:\(v_4\)] (v4) {};

\begin{pgfonlayer}{background}
\draw[edge,color=blue](v1)--(v3);
\draw[edge,color=blue](v2)--(v3);
\draw[edge,color=blue](v1)--(v4);
\draw[edge,color=blue](v2)--(v4);
\filldraw[fill=green](v3) circle (0.3cm);
\filldraw[fill=green](v4) circle (0.3cm);
\end{pgfonlayer}
\end{tikzpicture}
}
\subfigure[$\ket{10_L}$]
{
\begin{tikzpicture}
\node[vertex,label=below:\(v_1\)] (v1) {};
\node[vertex,right of=v1,label=below:\(v_2\)] (v2) {};
\node[vertex,above of=v1,label=above:\(v_3\)] (v3) {};
\node[vertex,right of=v3,label=above:\(v_4\)] (v4) {};

\begin{pgfonlayer}{background}
\draw[edge,color=blue](v1)--(v3);
\draw[edge,color=blue](v2)--(v3);
\draw[edge,color=blue](v1)--(v4);
\draw[edge,color=blue](v2)--(v4);
\filldraw[fill=green](v1) circle (0.3cm);
\filldraw[fill=green](v3) circle (0.3cm);
\filldraw[fill=green](v4) circle (0.3cm);
\end{pgfonlayer}
\end{tikzpicture}
}
\caption{The code $C_2$ constructed from the given set S}
\label{figEAECC-C2}
\end{figure}

The given choice of $S$ admits two possible codes. 
The first Code $C_1$ is given by the codewords $\ket{00_L}$ and $\ket{11_L}$ in Fig \ref{figEAECC-C2}. 
This code encodes one logical qubit and can correct all possible errors on the unsafe qubits 3 and 4. This can easily be checked by applying the KL-Condition with the error set consisting of all possible Pauli operators on qubits 3 and 4.
Note that this code could also have been realized by adding just one new vertex to $D$.

The second Code $C_2$ is given by all the codewords in Fig \ref{figEAECC-C2}.
This code encodes two logical qubits. However, it can only correct all single qubit errors on the two unsafe qubits, so the error correction capabilities are weaker than that of the code $C_1$.

\section{Conclusion}

In this report, error correction codes consisting of symmetric hypergraph states were related to binomial coefficients and in that way it was possible to construct all codes of distance 2 for up to 30 qubits. It was confirmed that no symmetric hypergraph code for up to 30 qubits has a distance of 3 or more.
It was proved analytically that no symmetric graph code with a distance of 3 or more exists for any number of qubits.
In addition, some hypergraph codes using protected qubits were constructed.

For further investigation it would be interesting to know if there exist symmetric hypergraph codes with distance higher than 2. The method could either be analytical examination or a continuation of the computer search to a higher number of qubits.
Furthermore, error correction with non-symmetric hypergraph codes could be explored, as it its possible that non-symmetric hypergraph codes will perform better than the symmetrical codes.
Another question is how the decoding of hypergraph codes using protected qubits can be implemented, and whether these codes can offer any advantages over the existing graph codes introduced in \cite{EAQECC}.

\section*{References}

\bibliography{bib}{}
\bibliographystyle{unsrt}

\section*{Appendix: Proofs}

\textbf{Proof of lemma \ref{ecc}:}

First note that, because the states are symmetric, the condition that $\Pi_{i=1}^{l-1}Z_i\ket{D}$ is balanced is equivalent to the condition that $\Pi_{i=2}^{l}Z_i\ket{D}$ is balanced, because only the number, not the position of the Z gates matters. In the following, this equivalent condition will be used for convenience of calculation.

Let $E_i$ be a Pauli $X$, $Z$ or $XZ$ operator acting on a single qubit $i$.
It follows directly from the KL-condition that the code C has
distance at least 2 if and only if the diagonal condition:
$\braket{H|E_i|H} = \braket{H^\prime|E_i|H^\prime}$
and the off-diagonal condition:
$\braket{H|E_i|H^\prime} = 0$
are fulfilled.
\\

"if direction":

Let $i \in \{1,..,n\}$.

From equation (\ref{Xdifference}) and Lemma \ref{ortho} it follows:

\begin{equation}
\label{eq14}
\braket{H|X_i|H} = \bra{+}^{\otimes n} (\ket{H} \Delta X_i\ket{H}) = \bra{+}^{\otimes n}\ket{+}_i\ket{D}_{\overline{i}} = \bra{+}^{\otimes n-1}\ket{D}_{\overline{i}} = 0
\end{equation}
where the last equality holds because $\ket{D}$ is balanced.
Now we can also compute:
\begin{equation}
\braket{H^{\prime}|X_i|H^{\prime}} = \braket{H|(\Pi_{j=1}^l Z_j) X_i (\Pi_{j=1}^l Z_j)|H} = \pm \braket{H|X_i|H} = 0
\end{equation}
The $\pm$ appears because $X_i$ might anti-commute with $\Pi_{j=1}^l Z_j$.
So we see that the diagonal condition holds for the operator  $X_i$ for any $i \in \{1,..,n\}$.

Now the off-diagonal condition is shown for the operator $X_1$:

\begin{eqnarray}
\braket{H|X_1|H^\prime} &= \braket{H|X_1\Pi_{j=1}^l Z_j|H} = -\bra{+}^{\otimes n}\Pi_{j=1}^l Z_j\ket{+}_1\ket{D}_{\overline{1}} \nonumber \\
&= -\bra{+}^{\otimes n}\Pi_{j=2}^l Z_j\ket{-}_1\ket{D}_{\overline{1}} = 0
\end{eqnarray}

Here, the second equality follows from equation (\ref{Xdifference}), lemma \ref{ortho}, the fact that $X$ and $Z$ anti-commute, and the fact that $Z$ and $C_e$ commute for any generalized controlled $Z$ gate $C_e$.
The last equality holds because $\braket{+|-} = 0$.
From this, it follows that the off-diagonal condition holds for all $X_i$ with $i \leq l$ because of the symmetry of the states.

For $i>l$: 

\begin{eqnarray}
\braket{H|X_i|H^\prime} &= \braket{H|X_i\Pi_{j=1}^l Z_j|H} \nonumber \\ &= \bra{+}^{\otimes n} \Pi_{j=1}^l Z_j \ket{+}_i\ket{D}_{\overline{i}} = \bra{+}^{\otimes n-1}\Pi_{j=1}^l Z_j\ket{D} = 0
\end{eqnarray}

The last equality holds because $\Pi_{i=1}^l Z_i\ket{D}$ is balanced.

This shows the diagonal and the off-diagonal conditions for all possible $X$ operators.

For the operator $X_iZ_i$ ($i \in \{1,..,n\}$) it holds, with an analogous calculation to equation (\ref{eq14}):
\begin{equation}
\braket{H|X_iZ_i|H} = \bra{+}^{\otimes n}\ket{-}_i\ket{D}_{\overline{i}} = 0
\end{equation}
and 
\begin{equation}
\braket{H^{\prime}|X_iZ_i|H^{\prime}} = \braket{H|\Pi_{j=1}^l Z_j X_iZ_i \Pi_{i=j}^l Z_j|H} = \pm \braket{H|X_iZ_i|H} = 0
\end{equation}

So the diagonal condition holds for all $X_iZ_i$ with $i \in \{1,..,n\}$.

For the off-diagonal condition:
\begin{eqnarray}
\braket{H|X_1Z_1|H^\prime} &= \braket{H|X_1Z_1\Pi_{i=1}^l Z_i|H} = -\bra{+}^{\otimes n}Z_1\Pi_{i=1}^l Z_i\ket{+}_1\ket{D}_{\overline{1}} \nonumber \\
&= -\bra{+}^{\otimes n}\Pi_{i=2}^l Z_i\ket{+}_1\ket{D}_{\overline{1}} = 0
\end{eqnarray}

where the last equality follows from the fact that  $\Pi_{i=2}^{l}Z_i\ket{D}$ is balanced. With this, the off-diagonal condition is shown for all $i<l$ because of symmetry. 

For $i>l$:
\begin{equation}
\braket{H|X_iZ_i|H^\prime} = \braket{H|X_iZ_i\Pi_{j=1}^l Z_j|H} = \bra{+}^{\otimes n}\Pi_{j=1}^l Z_j\ket{-}_i\ket{D}_{\overline{i}} = 0
\end{equation}

Now the conditions have been shown for all $X_iZ_i$.

The last case is the operator $Z_i$, but as the hypergraph basis is closed under this operation the calculations are clear, and the conditions hold because $l>1$. More specific, all the scalar products that have to be calculated will evaluate to 0, and the two conditions hold.
\\

"Only if direction":

Here, some of the calculations above will be used to show the other direction of the equivalence. 

Assume that the KL-condition holds for the given code.
It holds:

\begin{equation}
\braket{H^{\prime}|X_1|H^{\prime}} = \braket{H|\Pi_{i=1}^l Z_i X_1 \Pi_{i=1}^l Z_i|H} = - \braket{H|X_1|H}
\end{equation}
(using $l>1$)

On the other hand, from the diagonal condition:
\begin{equation}
\braket{H^{\prime}|X_1|H^{\prime}} = \braket{H|X_1|H}
\end{equation}

This implies $\braket{H|X_1|H} = 0$.
Therefore: $\braket{H|X_1|H} = \bra{+}^{\otimes n-1}\ket{D} = 0$. It follows that $\ket{D}$ must be balanced. 

Furthermore, for $i>l$:
\begin{equation}
\braket{H|X_i|H^\prime} = \braket{H|X_i\Pi_{j=1}^l Z_j|H} = \bra{+}^{\otimes n-1}\Pi_{j=1}^l Z_j\ket{D} = 0
\end{equation}
where the last equality holds because of the off-diagonal condition.
It immediately follows that $\Pi_{i=1}^l Z_i\ket{D}$ must be balanced.

Finally:
\begin{eqnarray}
\braket{H|X_1Z_1|H^\prime} &= \braket{H|X_1Z_1\Pi_{i=1}^l Z_i|H} = -\bra{+}^{\otimes n}Z_1\Pi_{i=1}^l Z_i\ket{+}_1\ket{D}_{\overline{1}} \nonumber \\
&= -\bra{+}^{\otimes n}\Pi_{i=2}^l Z_i\ket{+}_1\ket{D}_{\overline{1}} = -\bra{+}^{\otimes n-1}\Pi_{i=2}^l Z_i\ket{D} \nonumber \\
 &=  0
\end{eqnarray}
where the last equality again follows from the off-diagonal condition. It follows that 
$\Pi_{i=2}^l Z_i\ket{D}$ must be balanced.

\begin{flushright}
$\square$
\end{flushright}

\textbf{Proof of lemma \ref{eccBinomial}}:

Let $C = span(\ket{H},\ket{H^\prime})$, Let $\ket{D} = \ket{H^{m_1,..,m_k}_{n}}$ and let $M = \{m_1,...,m_k\}$.

The conditions in lemma \ref{ecc} are rewritten here:

Condition (i):

A state is balanced if half of the coefficients are negative. The number of negative coefficients is $\sum_{m \in M} {n \choose m} = 2^{n-1}$ for a symmetric state of this form. (Note that the number of computational basis states with weight $m$ on $n$ qubits is ${n \choose m}$)

Condition (ii):

This follows from condition (i) and the following statement:

If $\ket{H^{m_1,..,m_k}_{n}}$ is balanced, then the state $\Pi_{i=1}^l Z_i\ket{H^{m_1,..,m_k}_{n}}$ is balanced if and only if $\sum_{m \in M}\sum_{j \ \rm{odd}, j \leq l, j \leq m} {l \choose j}{n-l \choose m-j} = 2^{n-2}$.

Proof:

Applying the operator  $\Pi_{i=1}^l Z_i$ changes the sign of a computational basis state $\ket{s}$ if the number of ones contained in the first $l$ bits of the bit-string $s$ is odd.
As the state $\ket{H^{m_1,..,m_k}_{n}}$ is assumed to be balanced, $\Pi_{i=1}^l Z_i\ket{H^{m_1,..,m_k}_{n}}$ can only be balanced if applying $\Pi_{i=1}^l Z_i$ changes as many coefficients from negative to positive as it changes coefficients from positive to negative so the total number of negative coefficients remains unchanged.
The total number of sign changes is:
\begin{equation*}
2^{(l-1)}\cdot2^{(n-l)} = 2^{n-1}
\end{equation*}
because there must be an odd number of ones in the first $l$ qubits and the remaining $n-l$ qubits can have any state. 
This means that the number of sign changes from negative to positive has to be $2^{n-2}$. Let $t$ be the number of states with a negative coefficient that contain an odd number of ones in the first $l$ qubits. Then it follows $t = 2^{n-2}$.

The states with negative coefficients are the states with a weight in $M$.

Therefore this number is:

\begin{equation}
t = \sum_{m \in M}\sum_{j \ \rm{odd}, j \leq l, j \leq m} {l \choose j}{n-l \choose m-j} = 2^{n-2}
\end{equation}

Here ${l \choose j}$ is the number of possibilities of choosing $j$ ones in the first $l$ qubits, and ${n-l \choose m-j}$ is the number of possibilities to choose the remaining $m-j$ ones in the other $n-l$ qubits.

Condition (iii):

If $\ket{H^{m_1,..,m_k}_{n}}$ is balanced, then the state $\Pi_{i=2}^l Z_i\ket{H^{m_1,..,m_k}_{n}}$ is balanced if and only if:
$ \sum_{m \in M}\sum_{j \ \rm{odd}, j \leq l-1, j \leq m} {l-1 \choose j}{n-(l-1) \choose m-j} = 2^{n-2}$

Proof:

The proof is the same as for condition (ii).

This concludes the proof of lemma \ref{eccBinomial}.

\begin{flushright}
$\square$
\end{flushright}

\textbf{Proof of lemma \ref{lemmaGeneralEAECC}:}

%Consider an operator $E = E_iE_j^\dagger$ with $E_i,E_j \in R$.

First the KL-condition is considered for  two codewords arising from the same set $I \in S$.

It is  sufficient to show that for all subsets $x, z \subset A$ ($x$ and $z$ can overlap) with $|x \cup z| < |A|$ it holds:
\begin{equation}
\braket{H_I|E|H_I} = \braket{H^\prime|E|H^\prime}
\end{equation}
and
\begin{equation}
\braket{H_I|E|H_I^\prime} = 0
\end{equation}
where $E = \Pi_{i \in x} X_i \Pi_{j \in z} Z_j$.

First, consider the case that $|x|$ is even. Then, $\Pi_{i \in A} Z_i$ commutes with $E$ and therefore it follows:
\begin{equation}
\braket{H_I^\prime|E|H_I^\prime} = \braket{H_I|\Pi_{i \in A} Z_i \ E \ \Pi_{i \in A} Z_i|H_I} =\braket{H_I|E|H_I}
\end{equation}

Furthermore, because each operator $X_i$ has the same effect on $\ket{H}$ and $|x|$ is even, it holds: 
\begin{equation}
\Pi_{i \in x} X_i \ket{H_I} = \ket{H_I}
\label{HEH=1}
\end{equation}

Consequently:
\begin{eqnarray}
\braket{H_I|E|H_I^\prime} &= \braket{H_I|\Pi_{i \in x} X_i \Pi_{j \in z} Z_j|H_I^\prime} = \braket{H_I|\Pi_{j \in z} Z_j|H_I^\prime} \nonumber \\
&= \braket{H_I|\Pi_{j \in z} Z_j \Pi_{i \in A} Z_i|H_I} = 0
\end{eqnarray}

where the last equality follows because $|z| < |A|$, so at least one $Z$ gate remains.

This concludes the case that $|x|$ is even.

Now consider the case that $|x|$ is odd. Then for $j \in x$:
\begin{equation}
\Pi_{i \in x} X_i \ket{H_I} = X_j \Pi_{i \in x \setminus j} X_i \ket{H_I} =  X_j \ket{H_I}
\end{equation}
because $|x\setminus j|$ is even.

Then, by construction:
\begin{equation}
\braket{H_I|\Pi_{i \in x} X_i|H_I} = \braket{H_I|X_j|H_I} = \bra{+}^{\otimes n + e}\ket{+}^{\otimes n}_A\ket{D}_P = 0
\end{equation} 

Therefore:
\begin{eqnarray}
\braket{H_I|E|H_I} &= \braket{H_I|\Pi_{i \in x} X_i \Pi_{j \in z} Z_j|H_I} = \pm \braket{H_I| \Pi_{j \in z} Z_j \Pi_{i \in x} X_i|H_I} \nonumber \\
&= \pm \bra{+}^{\otimes n + e}\Pi_{j \in z} Z_j\ket{+}^{\otimes n}_A\ket{D}_P = 0
\end{eqnarray}
The last equality still holds because the $Z$ gates only operate on $A$.

In the same way it follows:
\begin{equation}
\braket{H_I^\prime|E|H_I^\prime} = 0
\end{equation}

Finally, with an analogous calculation:
\begin{equation}
\braket{H_I|E|H_I^\prime} = \pm \bra{+}^{\otimes n + e}\Pi_{i \in A} Z_i\Pi_{j \in z} Z_j\ket{+}^{\otimes n}_A\ket{D}_P = 0
\end{equation}
where the last equality follows again because all $Z$ gates act only on $A$.

Now, the case of codewords arising from two different sets $I,J \in S$ is considered.

For the condition $\ket{H_I|E\ket{H_J}} = 0$ ($I,J \in S$ with $I \neq J$) the following calculation is useful:

It is possible to expand $\ket{D}$ in the hypergraph basis of $\ket{+}^{\otimes e}$:
\begin{equation}
\ket{D} = \sum_{K \subseteq \{1,..,e\}} a_K \Pi_{i \in K} Z_i \ket{+}^{\otimes e}
\end{equation}

By the definition of $O$ it holds: $a_K = 0$ for all $K \in O$.

Therefore, it follows that $\bra{+}^{\otimes e}| \Pi_{i \in K} Z_i \ket{D} = 0$ for all $K \in O$,

meaning that $\Pi_{i \in K} Z_i \ket{D}$ is balanced for all $K \in O$.

Without loss of generality, it can be assumed that $E$ is of the form:

$E = \Pi_{i \in x} X_i \Pi_{j \in z} Z_j$ with $x,z \subset A$ and $|x \cup z| < |A|$. 

Using arguments similar to those above, it is calculated for the case that $|x|$ is even:

\begin{eqnarray}
\braket{H_I|E|H_J} = \braket{H_I|\Pi_{i \in x} X_i \Pi_{k \in z} Z_k|H_J} = \braket{H_I|\Pi_{k \in z} Z_k|H_J} = 0
\end{eqnarray}
 where the last equality follows because the $\Pi_{k \in z} Z_k$ acts only on the qubits in $A$.

For the case that $|x|$ is odd, again with similar arguments:

\begin{eqnarray}
\braket{H_I|E|H_J} &= \braket{H_I|\Pi_{i \in x} X_i \Pi_{k \in z} Z_k|H_J} \nonumber \\ 
&= \pm \braket{H|\Pi_{i \in I} Z_i\Pi_{k \in z} Z_k\Pi_{j \in J} Z_j\Pi_{i \in x} X_i |H} \nonumber \\ 
&= \pm \bra{+}^{\otimes (n+e)} \Pi_{k \in z} Z_k \Pi_{i \in (I \Delta J)} Z_i \ket{D}_P\ket{+}_A = 0
\end{eqnarray}

where the last equality follows because $\Pi_{k \in z} Z_k$ acts only on $A$ and $(I \Delta J) \in O$ by assumption.

Finally, the condition $\bra{H_I} E \ket{H_J^\prime} = 0$ must be considered. However, the calculations are the same as before.

\begin{flushright}
$\square$
\end{flushright}

\end{document}